\documentclass{aa}

\newcommand{\FO}{{\em FORS2}}
\newcommand{\FOR}{{\em FORS1}}
\newcommand{\VLT}{{\em VLT--UT2} }
\newcommand{\VLTT}{{\em VLT--UT1} }
\newcommand{\E}{{\it 3.6m ESO} }
\newcommand{\DOL}{{\em DoLoReS}}
\newcommand{\EF}{{\em EFOSC2} }

\newcommand{\src}{{RX\,J0806.3+1527}}
\newcommand{\srchri}{1BMW J080622.8+152732}

\newcommand{\bc}{\begin{center}}
\newcommand{\ec}{\end{center}}
\newcommand{\be}{\begin{equation}}
\newcommand{\ee}{\end{equation}}
\input psfig.sty
\newcommand {\rc}{\rm}
\begin{document}

\title{\src: a double degenerate binary with the shortest known orbital 
period (321s)\thanks{Based on observations carried 
out at ESO VLT (60.A--9203 and 66.D--0675) and La Silla (64.H--0604).}}  

\author{G.L. Israel$^{1,}$\thanks{Affiliated to I.C.R.A.}, W. 
Hummel$^2$, S. Covino$^3$, 
S. Campana$^{3,\star\star}$,  I. Appenzeller$^4$, W. G\"assler$^{5,6}$, 
K.-H. Mantel$^5$, G. Marconi$^{7,1}$, C.W. Mauche$^8$, 
U. Munari$^9$, I. Negueruela$^{10}$, H. Nicklas$^{11}$, G. Rupprecht$^2$, 
R.L. Smart$^{12}$, O. Stahl$^4$ and L. Stella$^{1,\star\star}$}

\institute{INAF -- Osservatorio Astronomico di Roma, V. Frascati 33, 
       I--00040 Monteporzio Catone, Italy
\and
European Southern Observatory, Karl--Schwarzschildstr. 2, D--85748 Garching, Germany
\and
INAF -- Osservatorio Astronomico di Brera, Via Bianchi 46, I--23807
Merate, Italy
\and
Landessternwarte Heidelberg, K\"onigstuhl 12, D--69117 Heidelberg,
Germany
\and 
Institut f\"ur Physik, 
Universit\"ats-Sternwarte M\"unchen, Scheinerstr. 1, D--81673
M\"unchen, Germany
\and
Subaru Telescope, NAOJ, 650 North A'ohoku Place, Hilo, Hawaii, HI 96720 USA
\and
European Southern Observatory, Casilla 19001, Santiago, Chile
\and
Lawrence Livermore National Laboratory, L--43, 7000 East Avenue,
Livermore, CA 94550 USA
\and
INAF -- Osservatorio Astronomico di Padova, Sede di Asiago, I--36012
Asiago, Italy
\and
Observatoire Astronomique de Strasbourg, rue de l'Universit\'e 11,
F67000 Strasbourg , France
\and
Universit\"atssternwarte G\"ottingen, Geismarlandstr. 11, D--37083
G\"ottingen, Germany
\and
INAF -- Osservatorio Astronomico di Torino, Strada Osservatorio 20,
I--10025, Pino Torinese (To), Italy
}

\date{Received: 12 February 2002; Accepted: 1 March 2002} 
\offprints{gianluca@mporzio.astro.it} 
\authorrunning{Israel et al.} 
\titlerunning{321\,s -- the shortest orbital period binary.}

\thispagestyle{empty}
\abstract{ 
We carried out optical observations of the field of the X--ray
pulsator \src. A blue V=21.1 star was found to be the only object
consistent with the X--ray position.  VLT FORS spectra revealed a blue
continuum with no intrinsic absorption lines. Broad ($v$$\sim$1500
km\,s$^{-1}$), low equivalent width ($\sim$--1$\div$--6\AA) emission
lines from the HeII Pickering series were clearly detected. B, V and R
time--resolved photometry revealed the presence of $\sim$15\%
pulsations at the $\sim$321\,s X--ray period, confirming the
identification. These findings, together with the period stability and
absence of any additional modulation in the 1\,min$ - $5\,hr period
range, argue in favour of the orbital interpretation of the 321\,s
pulsations. The most likely scenario is thus that \src\ is a double
degenerate system of the AM CVn class. This would make \src\ the
shortest orbital period binary currently known and one of the best
candidates for gravitational wave detection.
 }
\maketitle
\section{Introduction} 

During a systematic search for periodic signals in a sample of
$\sim$\,4000 ROSAT HRI (0.1--2.4 keV) light curves we discovered
321.25\,s pulsations in the X--ray flux of \src\ (Israel et al. 1999,
thereafter I99). Based on the large pulsed fraction ($\sim$100\%),
relatively low 0.5--2.0 keV flux (3.0--5.0\,$\times$\,10$^{-12}$ erg
cm$^{-2}$\,s$^{-1}$), modest distance (edge of the Galaxy is at
$\leq$1\,kpc in the direction of the source) and presence of a faint
(B=20.7) blue object in the Digitized Sky Survey 1\farcs5 away from
the X--ray position, the source was tentatively classified as a
cataclysmic variable of the intermediate polar class (I99).  A similar
classification was also suggested by Beuermann et al. (1999; see also
Burwitz \& Reinsch 2001) based on the source X--ray colours in the
ROSAT all sky survey (RASS).
\begin{figure}[bth]
\centerline{\psfig{figure=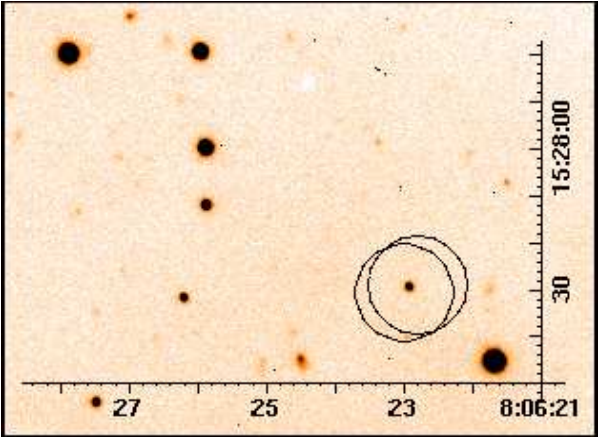,width=7.9cm}}
\caption{VLT FORS2 B filter image of the field of \src. The circles 
represents the X--ray uncertainties inferred from the ROSAT HRI 
observations. Coordinates units are right ascension and declination.}
\end{figure}
\begin{table*}[tbh]
\begin{center}
\caption{Optical observations carried out in 1999--2001 for \src}
\begin{tabular}{ccccll}
\hline 
Obs.  & Telescope \& Instrument& Date & Exp.  & Range  & Comments\\
 \#  &                   &       &  (s)  &(Band/\AA) &  \\ 
\hline 
A &8.2m \VLT\  \FO   &1999/Nov/12 &300&BVRIH$_{\alpha}$H$_{\alpha {\rm cont}}$&  \\
B & ''   &1999/Nov/16--19 &10500&3600--6000& 6\AA\ res.; grism 600B; slit 1\arcsec\\
C & ''   &1999/Nov/18 &600$\div$300&UBVRI& \\
D & ''   &2000/Feb/05&2700&3600--9000 & 30\AA\ res.; grism 150I; slit 1\arcsec\\
\hline
E &\E\  \EF  & 2000/Mar/30&320&UBVRI&\\
F & ''  & '' &1800&B& TRP; exposures of 20\,s each\\
\hline 
G & 3.6m TNG \DOL & 2001/Jan/01 & 21600 & BVR & TRP; exposures of 20\,s each \\
\hline 
H & 8.2m \VLTT\  \FOR & 2001/Jan/16--23 & 18000 & 3600--6000 & 6\AA\ res.; grism 600B; slit 1\arcsec\\
I & '' & 2001/Jan/19--24 & 10800 & 6000--9000 & 30\AA\ res.; grism 150I; slit 1\arcsec \\
  \hline 
\end{tabular}  
\end{center}
\end{table*}

In this letter we report the results of optical campaigns carried out
in 1999--2001, which confirm unambiguously the identification of the
optical counterpart of \src\ suggested by I99 and
provide evidence that \src\ is a double white dwarf interacting binary
with the shortest known orbital period.
    
\section{Observations}
Table\,1 gives the journal of observations.  Several images in U, B,
V, R and I Bessel filters were taken together with three different
H$\alpha$ interference filters ($\Delta\lambda = 65$\AA;
obs. A--E). Standard procedures were used for bias subtraction and
flat--field correction. Figure\,1 shows the B filter image of the
field of \src\ together with the ROSAT HRI positional
uncertainties. Down to the limiting R magnitude of $\sim$25.3 only one
object was found consistent with the HRI error circles. This was at
R.A.\,=\,08$^{\rm h}$ 06$^{\rm m}$~22\fs9 and Dec.\,=\,+15$^{\circ}$
27\arcmin\ ~31\farcs0 (uncertainty of $\sim$0\farcs5; equinox 2000),
fully consistent with the previously identified blue object (I99).
The U, B, V, R and I magnitudes were determined to be 19.6, 20.7,
21.1, 21.0 and 20.9, respectively (magnitude uncertainties of
$\sim$0.1) on March 1999. 

Time--resolved photometry (TRP) in the B band was first obtained with
the ESO 3.6\,m telescope equipped with EFOSC2 (obs. F) over an
interval of 1800\,s. Pulsations at the 321\,s X--ray period were
detected, with a pulsed fraction (semiamplitude of modulation divided
by the mean source magnitude) of 15$\pm$4\% (90\% confidence level)
confirming the correctness of the optical counterpart identification.
  
Based on this result, we observed again the source in the B, V and R bands
with the 3.6\,m Telescopio Nazionale Galileo (TNG) equipped with
\DOL\ (obs. G). {\rc TRP at a time resolution of 20\,s was obtained in each 
of the three filters ($\sim$2\,hr per filter) for a total duration of
7\,hr}. Differential light curves were accumulated by subtracting the
source magnitudes with the {\rc mean} magnitude of 10 reference stars
within the field of view. Figure\,2 (upper panel) shows the whole
light curve of \src\ {\rc normalised, as an example, to one} of the
reference stars. A best period of 321.5$\pm$0.3\,s was obtained by
fitting the phases of the modulation obtained over 6 different
intervals of $\sim$\,4000\,s exposure each. This value is fully
consistent with that detected in X--rays (321.25$\pm$0.25\,s), to
within $\sim$1 part in 10$^3$.  The shape of the B, V and R
modulations could be well fit by the sum of two sinusoids (fundamental
plus 2nd harmonic; see Figure\,2). Pulsed fractions of 13.9$\pm$0.5\%,
14.2$\pm$0.6\% and 13.2$\pm$0.6\% (90\% confidence level) were
determined for the B, V and R band, respectively.  In order to search
for additional flux modulations up to periods of hours, we merged the
B, V and R light curves, by normalising their average flux to the
average B band flux and calculated a power spectrum (see Figure\,2);
no significant signal was detected (99\% confidence level upper limit
of 1.5\%) in the 1\,min$ - $5\,hr period range, other than that at
321\,s. Aperiodic flickering is apparent in the optical light curve. 

\begin{figure*}[bht]
\centerline{\psfig{figure=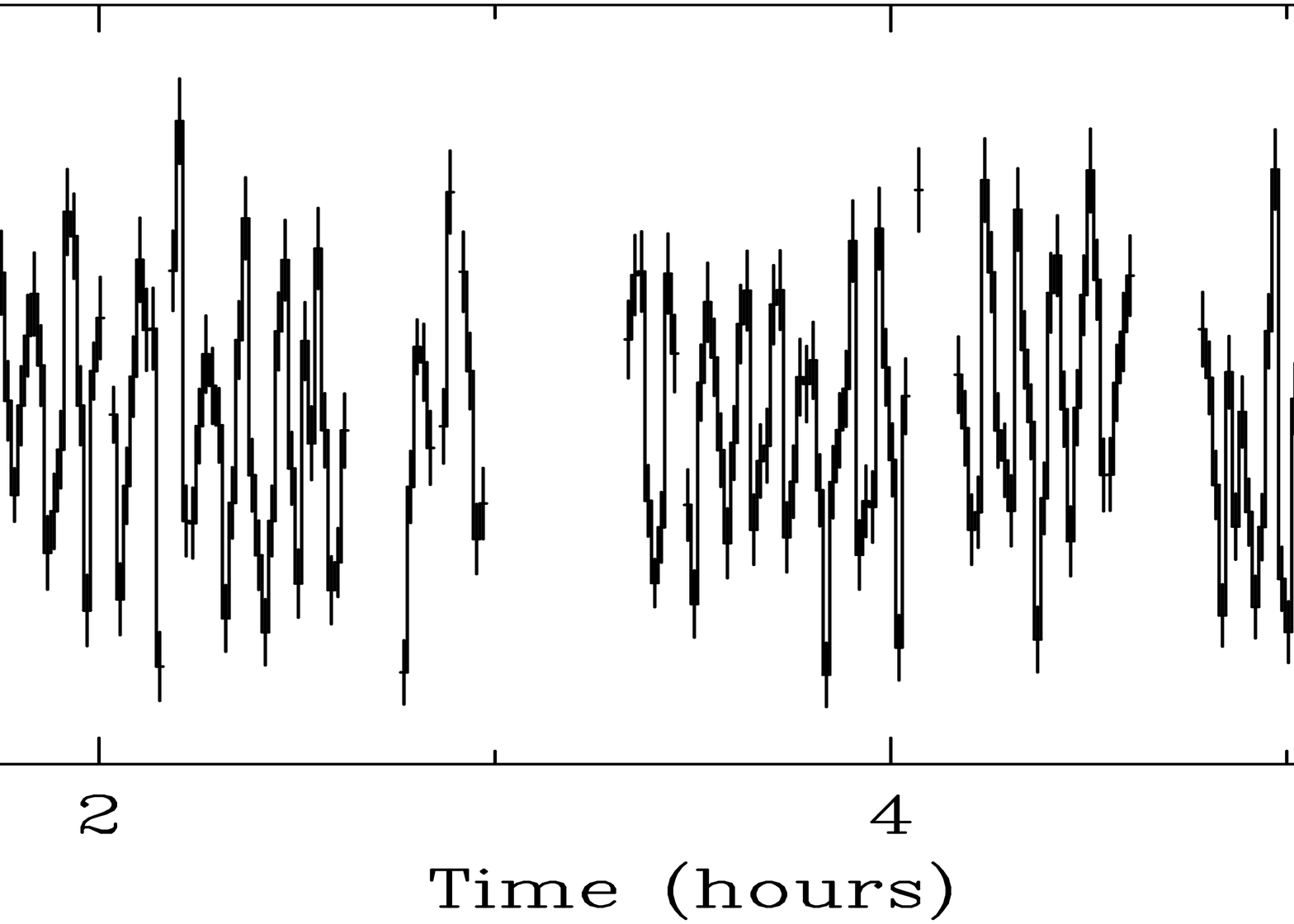,height=3.9cm,angle=-90}}
\centerline{\psfig{figure=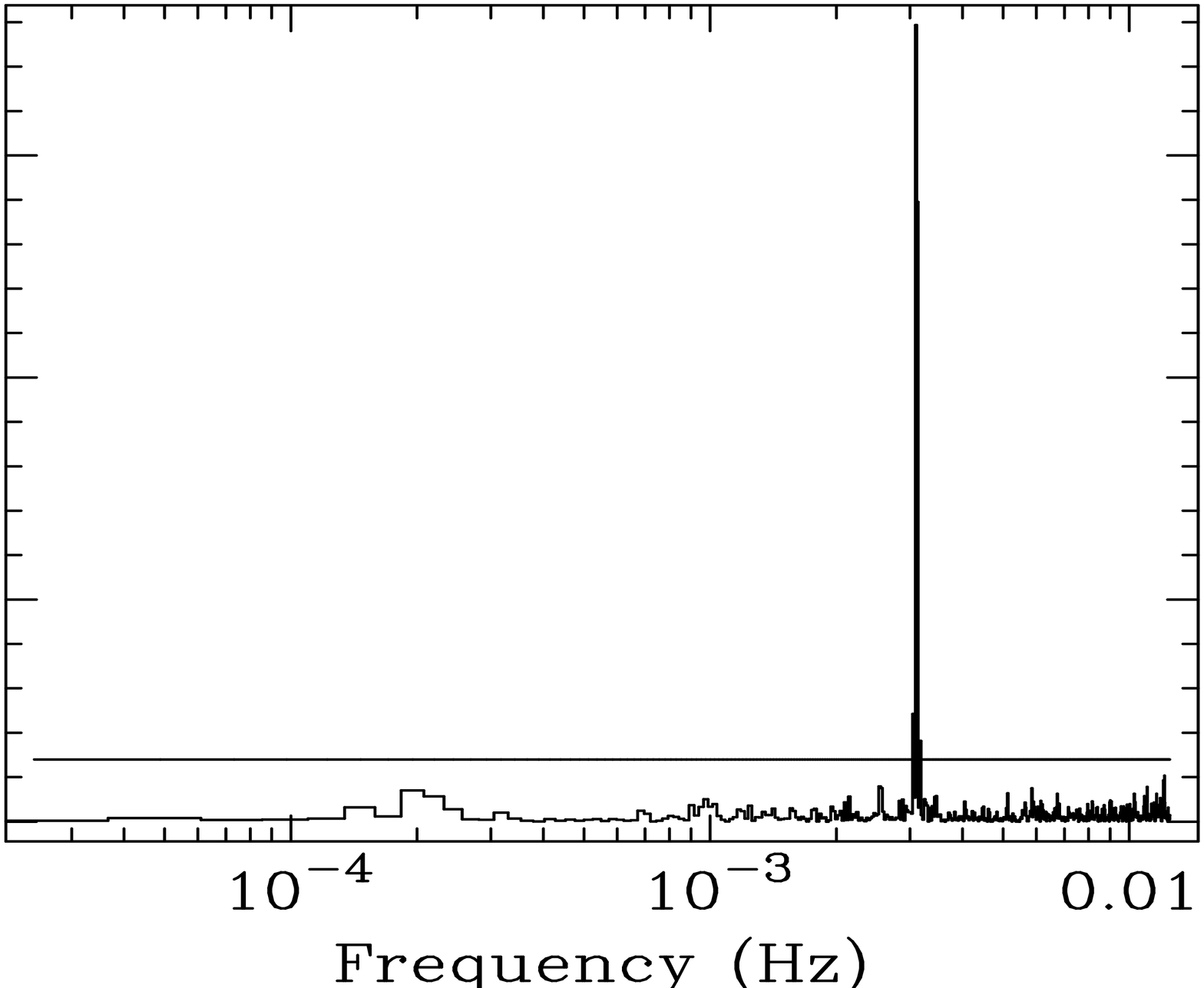,height=4.3cm,angle=-90}
\psfig{figure=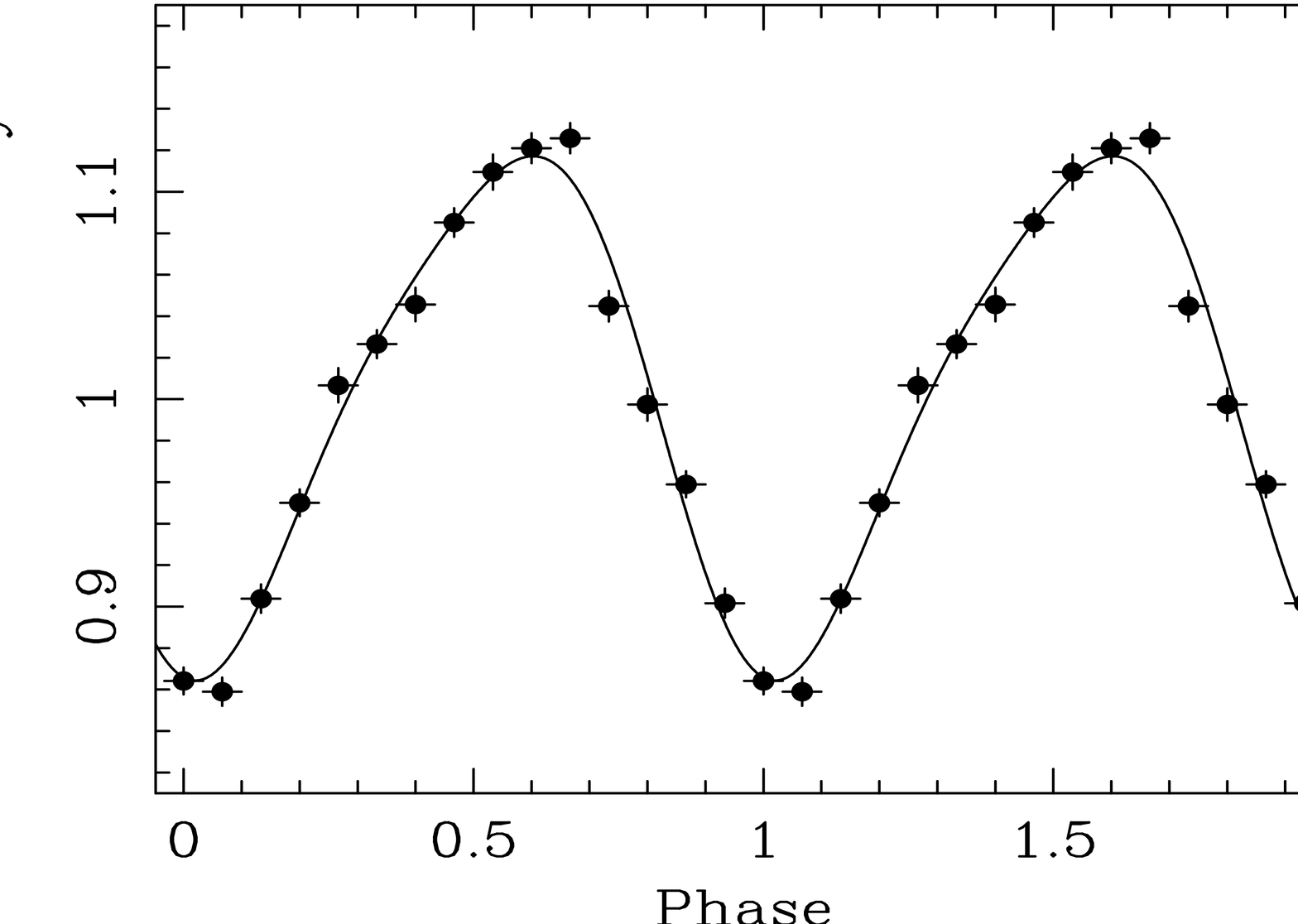,height=4.3cm,angle=-90}}
\caption{TNG \DOL\ B (from start to 2.1hr), V (2.1--4.5hr) and R (4.5hr to 
the end) merged light curve for the optical 
counterpart of \src\ (upper panel). Power spectrum density with
superimposed the 99\% confidence level threshold for signals (lower
left panel). Merged B, V and R light curve folded to the best period
of 321.5\,s, with superimposed the best fit (lower right panel).}
\end{figure*}

Several ESO Very Large Telescope (VLT) medium resolution (6\AA)
spectra were obtained with 30\,min exposures in the blue band with
FORS1+2 (see Rupprecht \& B\"ohnhardt 2000 for instrument description;
obs. B and H). Each spectrum was analysed independently. The summed
and normalised spectrum obtained during obs. H is shown in Figure\,3
(bias subtracted, flat--field corrected and calibrated in
wavelength). No significant absorption features were detected, while
several faint (equivalent width, EW, of $\sim$--1$\div$--6\AA) and
broad (full width half maximum, FWHM, of $\sim$20--30\AA) emission
lines are apparent.  The lines at 5411\AA, 4541\AA, 4199\AA, 4025\AA\
and 3923\AA\ are unambiguously identified with HeII Pickering
lines. The lines at 6560\AA, 4859\AA, 4338\AA, 4100\AA\ and 3968\AA\
correspond to the even terms of this series. {\rc NIII/CIII emission
lines around 4640\AA\ and 5270\AA\ are also detected testifying that
recombination processes are occurring in the system.} Intensity
changes (up to a factor of $\sim$2 in EW) and shape of the emission
lines were detected from night to night and even within the same
night.
\begin{figure*}[t]
\centerline{\psfig{figure=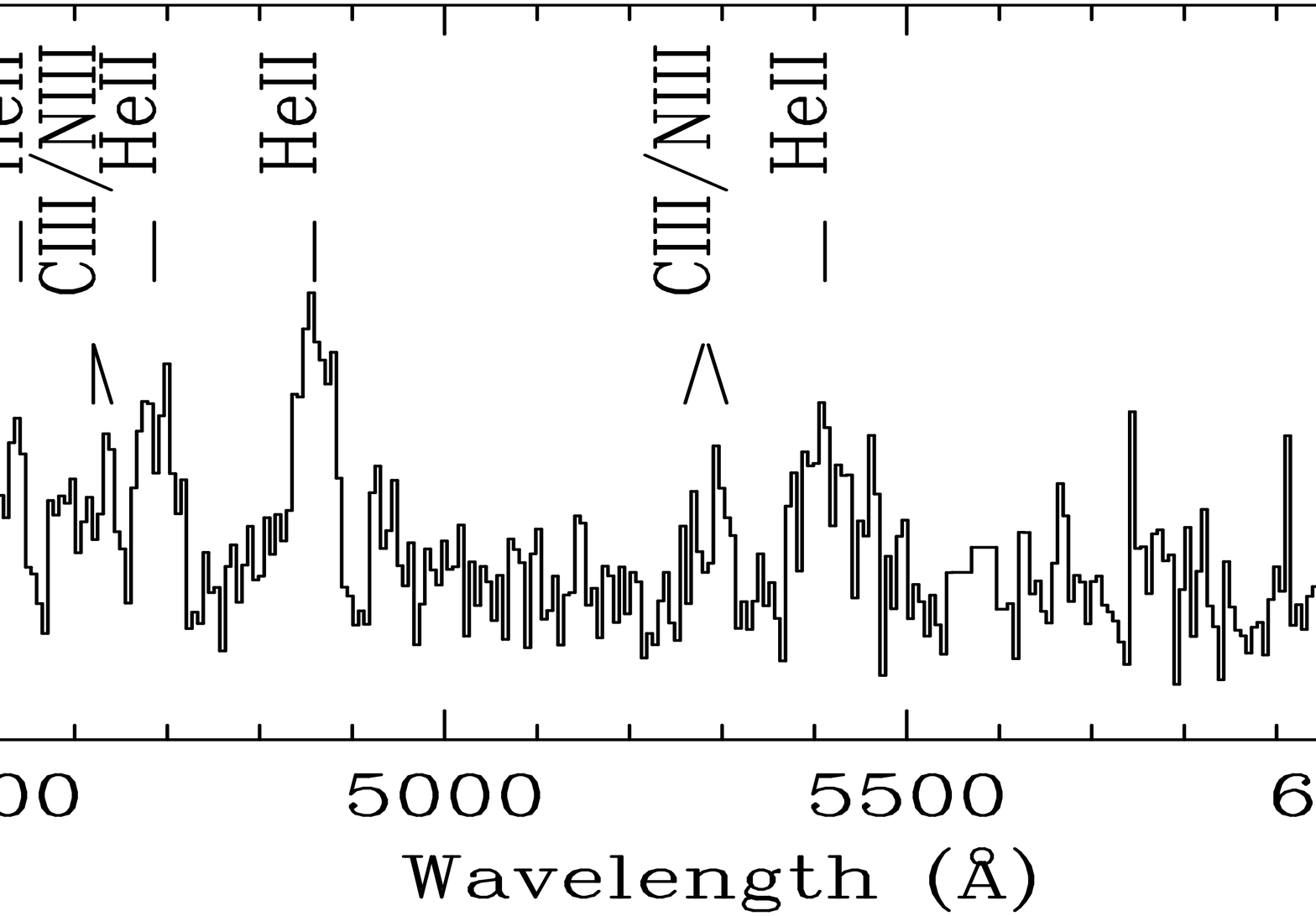,height=4.3cm,angle=-90}}
\caption{VLT FORS1 medium (6\AA; 3900--6000\AA) and low (30\AA; above 
6000\AA) resolution spectra obtained for the optical counterpart of
\src. Numerous faint emission lines of HeI and HeII (blended 
with H) are labeled.}
\end{figure*}
VLT low resolution (30\AA) spectra were also obtained to sample the
red part of the spectrum (obs. D and I; exposure times of
15--30\,min).  The low and medium resolution spectra collected in 2001
were flux--calibrated using the spectrum of the V=13.56 sdB {\rc
spectrophotometric} standard GD108. The calibrated flux is dominated
by a steep blue continuum with no intrinsic absorption lines (see
Figure\,4).  The shape of the optical continuum is thermal and implies
a temperature of T$_{\rm bb}$$>$4$\times$10$^4$\,K together with
negligible reddening of E(B--V)$\leq$ 0.01.  Its normalisation is such
that its blackbody emitting radius is $R_{\rm bb} \sim 570
(d/100$\,pc$)(T_{\rm bb}/4\times 10^4$\,K$)^{-1/2}$\,km, with $d$ the
source distance.

Spectral information at higher photon energies was obtained by
analysing the Extreme Ultraviolet Explorer (EUVE) survey and the RASS
data.  The EUVE upper limits were extrapolated from data taken from
on--line a strophysical databases. The relevant archival RASS data were
retrieved and photon arrival times extracted within a 2\arcmin\ radius
region centered on the peak of the emission. Photons were also
extracted from a nearby region, far from other detected sources, so as
to have a similar background level.  The source ROSAT PSPC Pulse Hight
Analyser (PHA) rates were grouped so as to contain a minimum of 10
photons per energy bin (after background subtraction), resulting in 4
statistically independent energy bins. A blackbody model gave a good
fit (by using a maximum likelihood and C statistics) with a
characteristic temperature of $T_{\rm
bb}\sim$(6$\pm^{13}_{4})\times10^{5}$\,K (90\% confidence level) and
an absorption column of 1.9$\times$10$^{20}$ cm$^{-2}$ (however the
90\% uncertainty includes all values $<$6 $\times$10$^{20}$ cm$^{-2}$)
and a 0.1--2.0\,keV unabsorbed flux of
$\sim$4$\times$10$^{-10}$\,erg\,cm$^{-2}$\,s$^{-1}$. We also inferred
an unabsorbed flux of
$\sim$5$\times$\,10$^{-9}$\,erg\,cm$^{-2}$\,s$^{-1}$ at the peak of
the emission of the 321\,s modulation. By combining the X--ray data
point and EUVE upper limits with the optical continuum from the VLT
observations, we obtained a best fit blackbody spectrum for a
temperature of T$\sim$2.6$\times$ 10$^5$\,K.  This value, however
should be treated with caution as it depends crucially on the
assumption that the spectrum of \src\ from the optical to the X--rays
is due to a single blackbody emission component. We note that the
vastly different amplitude of the 321\,s modulation in the optical and
the X--rays argues against this possibility.
\begin{figure}[bt]
\centerline{\psfig{figure=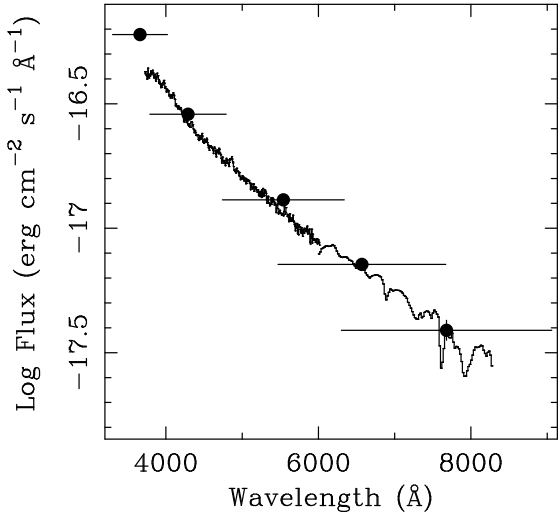,height=6.8cm,angle=-90}}
\caption{VLT FORS1 flux--calibrated medium and low resolution 
spectra with superimposed, from the left to the right, the U, B, V, R
and I photometric data.}
\end{figure}

Finally, we used the 1951 and 1991 Palomar Observatory Sky Survey
plates, digitized to produce the Guide Star Catalogue--II, in
conjunction with the ESO frames to put an upper limit of
0.02\arcsec\,yr$^{-1}$ on the proper motion of this object.  Assuming
that we are seeing only the reflex peculiar motion of the sun, this
would put the object at a distance of $\geq$100\,pc.
 
\section{Discussion}

The results of the optical campaigns reported in this letter led to
the unambiguous identification of the optical counterpart of
\src, and provide also important new constraints on the source. 

Firstly, the possibility that \src\ is a nearby isolated neutron star
accreting from the interstellar medium can be ruled out, because for a
distance of $\geq$100\,pc (as implied by the upper limit on the proper
motion) the blackbody emission radius inferred from the optical
continuum (or the optical continuum plus the RASS point) would be
several hundred km at least, too large a value for any neutron
star model.

Two additional results are especially relevant for assessing the
nature of \src; these are: (a) the absence of optical periodicities
other than the 321\,s modulation (up to periods of $\sim$5\,hr); (b)
the He emission line spectra.  The absence of a second periodicity
argues against models involving a non--synchronous magnetic accreting
white dwarf (such as the intermediate polars, IPs).  In addition to
their orbital and spin periodicities (and/or their beat), IPs display
optical spectra with intense Balmer emission lines with EW of tens of
\AA\ and widths of a few hundred km\,s$^{-1}$. Moreover the X--ray
amplitude of the spin modulation is comparatively low (Hellier
1999). The $100$\% X--ray modulation and broad and weak emission lines
that we observed from \src\ are also very much at variance with these
IP properties.  Moreover a hot blue thermal continuum similar to the
one we revealed from \src\ is simply not observed in magnetic
cataclysmic variables of any class.  We note that if the blue thermal
continuum were attributed to the accreting white dwarf (and/or the
accretion stream) the optical spectra exclude the presence of a
companion star earlier than a M7V for a (maximum) distance of 1\,kpc.
The condition that such a (hypothetical) companion star fills its
Roche lobe translates into an orbital period of $<$2\,hr, well within
the range of periods sampled by our photometric studies. Moreover it
would be very difficult the explain the observed He emission lines
within such a scenario.

The most natural interpretation of our results is that \src\ is a
double degenerate binary system, in which mass is transferred from a
Roche lobe filling white dwarf to another more massive white dwarf.
Such binaries belong to the AM CVn class and comprise only a few
objects, although recently two new members have been proposed
(RX\,J1914+24, period of 9.5\,min; Ramsay et al. 2000, and
KUV\,01584--0939, period of 10\,min; Warner \& Woudt 2002). AM CVns
are intrinsically blue objects with thermal optical continua, orbital
period in the 10--50\,min range (usually detected in the optical
band), fairly broad He emission and absorption lines
(FWHM$\sim$15--35\AA; for a review see Warner 1995). The presence of
flickering in the optical light curve testifies that these systems are
powered by accretion. AM CVns are expected to originate from normal
composition binaries which experienced two phases of mass exchange,
exposing the helium cores of the original stars (e.g. Warner 1995).
During their evolution, they reach a minimum orbital period of
$\sim$4\,min and thereafter evolve to longer periods. Double--peaked
emission line spectra clearly testify to the presence of an accretion
disk mediating the flow of matter. However in the X--ray bright AM CVn
candidate RX\,J1914+24, the accretion disk is likely not present
(Marsh \& Steeghs 2002; Wu et al. 2002). In one model for such a short
orbital period system the soft X--ray emission originates from direct
impact accretion which occur when the minimum distance of the gas
stream from the center of mass of the accretor is smaller than the
accretor size (Marsh \& Steeghs 2002).

We propose here that \src\ is a new member of the AM CVn class and in
particular that: (i) The modulation at 321\,s corresponds to the
orbital period of the system. The 100\% amplitude in the X--ray
modulation 
can be easily explained in terms of self--occultation of the stream
impact point on the surface of the accreting white dwarf.  The blue
optical continuum and the small amplitude optical modulation likely
results from X--ray reprocessing by a fraction of the donor star
surface and/or the accretion stream. The similarities with
RX\,J1914+24 suggest that \src\ might also be a direct impact
accretor. In this case the system may be non--synchronous, while no
conspicuous modulation is produced at the accreting white dwarf spin
period.  The possibility that the system is magnetically locked, as in
AM Her binaries, cannot be excluded at present, although the observed
line EWs and shape of the optical continuum argue against this.  (ii)
Emission lines from He are expected given the temperatures implied by
the optical continuum and the presence of H depleted accreting
gas. Note that He emission lines were detected also in
KUV\,01584--0939 (Wegner et al. 1987), whereas the study of the
optical spectrum of RX\,J1914+24 is hampered by the high absorption in
the direction of the source (Ramsay et al. 2000, {\rc 2002}).  The
presence of flickering in the \src\ optical light curves (and the
X-ray ROSAT HRI light curves as well, see I99) further suggests that
the optical emission is at least in part related to the accretion
process, as expected in the irradiation scenario.

The condition that the mass donor white dwarf fills its Roche--lobe
determines its mass M$_2$=0.12\,M$_{\odot}$, and radius
R$_2$=1.7$\times$10$^{9}$\,cm=0.02R$_{\odot}$. The mass transfer rate
\.M driven by angular momentum losses through gravitational radiation
can then be calculated over a reasonable accretor mass range.  We
adopt 0.2$\leq$M$_1$/M$_{\odot}$$\leq$0.5, the lower limit being close
to the stable mass transfer limit.  This range translates into
1--3$\times$10$^{-7}$M$_{\odot}yr^{-1}$ (assuming a efficient tidal
coupling), an accretion luminosity in the
2--5$\times$10$^{35}$\,erg\,s$^{-1}$ range and a flux of
10$^{-9}$--4$\times$10$^{-7}$\,erg\,s$^{-1}$\,cm$^{-2}$ (for
a distance in the 0.1--1\,kpc range). The latter values are
consistent with the peak flux inferred from the RASS (see Section 1).
In order not to violate the (minimum) size of the blackbody emitting
area derived from the normalisation of the optical continuum
(200$\leq$R$_{\rm bb}$(km)$\leq$6000 for a distance of 0.1--1\,kpc and
the entire range of allowed blackbody temperatures) the optical
continuum must come from a smaller region than the donor star, perhaps
the accretion stream and/or the illuminated part of the donor star.

Within this interpretation the secondary's orbital velocity is
expected to be in the 900--1500\,km\,s$^{-1}$ range, such that the
observed emission line width might be dominated by the binary's
Doppler velocity amplitude. Phase--resolved spectroscopy can ascertain
this point unambiguously and provide important new information.

In summary our results provide compelling evidence that \src\ is a
double white dwarf interacting binary system, with the shortest
orbital period ever recorded. The source represents one of the most
promising targets for gravitational wave detection from binary motion
(Nelemans et al.  2001). Indeed a 321\,s orbital period falls well
above $\log$$f$$\sim$--2.7\,Hz, where the average galactic background
hampers the gravitational wave detection. Moreover assuming reasonable
values of the distance and mass of the accretor, a strain amplitude of
up to few$\times 10^{-21}$ is expected, which is well above the LISA
sensitivity.

\begin{acknowledgements}
Based on observations made with the Italian Telescopio Nazionale
Galileo (TNG) operated on the island of La Palma by the Centro Galileo
Galilei of the CNAA (Consorzio Nazionale per l'Astronomia e
l'Astrofisica) at the Spanish Observatorio del Roque de los Muchachos
of the Instituto de Astrofisica de Canarias We thank the ESO director
for the DDT allocated to observe \src\ and R.M. Athreya for the help
in carrying out observations at the 3.6m ESO.  We thank G. Tessicini
and G. Marino for the observations at the TNG. The Guide Star
Catalogue--II is joint project of the Space Telescope Science
Institute and the Osservatorio Astronomico di Torino. {\rc An
anonymous referee provided useful comments on the first version of
this paper}. This work is supported through CNAA, ASI and MURST
grants.
\end{acknowledgements}


\end{document}